# Emergence of decoupled surface transport channels in bulk insulating Bi$_2$Se$_3$ thin films

Matthew Brahlek[1], Nikesh Koirala[1], Maryam Salehi[2], Namrata Bansal[3], and Seongshik Oh[1,*]

[1]Department of Physics & Astronomy, Rutgers, The State University of New Jersey, Piscataway, New Jersey 08854, U.S.A.

[2]Deparment of Materials Science and Engineering, Rutgers, The State University of New Jersey, Piscataway, New Jersey 08854, U.S.A.

[3]Department of Electrical and Computer Engineering, Rutgers, The State University of New Jersey, Piscataway, New Jersey 08854, U.S.A.

*Correspondence should be addressed to ohsean@physics.rutgers.edu

**Abstract:** **In ideal topological insulator (TI) films the bulk state, which is supposed to be insulating, should not provide any electric coupling between the two metallic surfaces. However, transport studies on existing TI films show that the topological states on opposite surfaces are electrically tied to each other at thicknesses far greater than the direct coupling limit where the surface wavefunctions overlap. Here, we show that as the conducting bulk channels are suppressed, the parasitic coupling effect diminishes and the decoupled surface channels emerge as expected for ideal TIs. In Bi$_2$Se$_3$ thin films with fully suppressed bulk states, the two surfaces, which are directly coupled below ~10 QL, become gradually isolated with increasing thickness and are completely decoupled beyond ~20 QL. On such a platform, it is now feasible to implement transport devices whose functionality relies on accessing the individual surface layers without any deleterious coupling effects.**

During the last several years, 3-dimensional (3D) topological insulators (TI) have garnered a lot of interest due to the exotic metallic states that are present on their surfaces[1-2]. Angle resolved photo-emission spectroscopy (ARPES)[3-5], scanning tunneling microscopy (STM)[6], and transport measurements[7-13] have shown that these topologically protected surface states (TSS) display massless Dirac-like linear dispersion with spin-momentum locking. However, existing TIs suffer from parasitic bulk conduction due to unavoidable self-doping effects. Progress has been made to reduce the bulk conductance both in bulk crystals utilizing anion mixing[14-15], and in thin films, where the large surface to volume ratio has made it relatively easy to achieve dominant surface transport properties[13]. However, when TI samples are made thin, the conducting channel provides an inter-surface conduction path, and this causes the top and bottom TSS to be coupled into a single channel in weak anti-localization (WAL) effect[12-13, 16-19]; it should be noted that this indirect coupling through the bulk state in the WAL effect occurs over a length scale far greater than the critical thickness (~6 quintuple layers, QL ≈ 1 nm) of direct coupling that has been observed in ARPES[20]. Further effort to suppress the bulk conduction in thin TIs led to a significant reduction in the mobilities, suppression of the surface quantum oscillations and strong inter-surface coupling[21-22]. Here, we report that by fully suppressing the parasitic channels in $Bi_2Se_3$ thin films through compensation doping, the TSS on opposite surfaces become fully decoupled from the bulk and each other with enhanced transport properties.

Although a number of valence 2+ elements have been shown to suppress the naturally occurring n-type carriers in bulk $Bi_2Se_3$ crystals, most of them were ineffective in thin films. Interestingly, copper, which has been used to induce superconductivity in $Bi_2Se_3$ [23], turns out to effectively suppress the n-type carriers in $Bi_2Se_3$ (see Ref. [24]) thin films without degrading their mobilities. In high mobility undoped $Bi_2Se_3$ films, the Hall resistance, $R_{xy}$, is typically non-linear due to multiple contributions from the TSS and other non-topological channels[13]. However, as shown in Fig. 1(a-b), as the Cu concentration ($x$ = Cu/Bi × 100%) was increased, $R_{xy}$ went from non-linear to linear at $x \approx 2.5 - 4.0\%$, and then became non-linear again beyond $x \approx 4.0\%$. This implies that at the optimal doping ($x \approx 2.5 - 4.0\%$), the transport is dominated by one type of carriers and thus $n_{Hall}$ can be extracted simply from $R_{xy} = $ $B/(en_{Hall})$ (these, and all measurements herein, were performed at 1.5K). For further discussion on possible doping mechanisms see Ref. [25].

To see the doping effect of Cu on $Bi_2Se_3$ films, Fig. 1(c) shows that the Hall carrier density, which exhibits a clear minimum at the doping where $R_{xy}$ became linear; the minimum Hall carrier density for these Cu-doped $Bi_2Se_3$ films is $n_{Hall} \approx 5.0 \times 10^{12}$ /cm$^2$, which is an order of magnitude lower than that of undoped $Bi_2Se_3$ ($n_{Hall} \approx 40 \times 10^{12}$ /cm$^2$) grown in otherwise identical conditions [13]. According to ARPES measurements on Cu-doped $Bi_2Se_3$[26], the band structure of TSS remained almost unchanged beyond these doping levels. As shown in Ref. [25], comparing the measured sheet carrier density with ARPES spectra [4, 27, 28] estimates a surface $E_F$ of ~146 meV above the Dirac point, which corresponds to ~70 meV below the bottom of the conduction band.

As can be seen in Fig. 1(d), the overall trend in the Hall mobility versus $x$ is monotonically downward (black solid line) with a more unusual spike in mobility near the optimal doping (red solid line). The downward trend is naturally expected with the increase in impurity scatting due to Cu dopants. In this regard, the upturn emphasized in red in Fig. 1(d) is surprising; one likely explanation is that Cu atoms do not act solely as chargeless defects, but rather Cu facilitates $Bi_2Se_3$ to grow with fewer Se vacancies near the critical regime, and thus Cu acts effectively as p-type dopant while enhancing the mobility.

Regarding $n_{Hall}$ vs. thickness (Fig. 1(e)), between ~10 – 150 QL, the Hall carrier density was nearly constant at $n_{Hall} \approx 6.0 \pm 2.0 \times 10^{12}$ /cm$^2$ (the slight increase in carrier density below 10 QL implies that Cu doping is less effective at small thicknesses.). This thickness-independent signature can originate either from a non-topological 2D electron gas (2DEG) due to quantum confinement or from the TSS. In order for a 2DEG to exist, downward band-bending is necessary at the surfaces[13], but with the estimated surface $E_F$, which is significantly below the conduction band minimum, if any band-bending is present it must be upward, giving rise to a depletion layer instead of an accumulation layer on the surfaces (see Ref. [25]); such upward band-bending near the surface cannot harbor a 2DEG. This naturally leads to the conclusion that the data is most consistent with the TSS as the only surface channel in the optimally-Cu-doped samples. Lastly, unlike the sheet carrier densities, the mobility versus thickness (Fig. 1(f)) does not show full thickness independence. Rather, at around 20 QL, there is an anomalous spike from ~1100 to 2500 cm$^2$/Vs. The exact origins of the mobility peaks observed in Fig. 1(d) and (f) are currently unclear and need further studies.

Near the optimal doping, Shubnikov-de Hass (SdH) oscillations were prominent enough to allow a standard Lifshitz-Kosevich (LK) analysis to be carried out on several samples. Undoped or non-optimally-doped Bi$_2$Se$_3$ samples were afflicted by multiple low-amplitude SdH channels, thus making LK analysis inapplicable [13]; as shown in Fig. 2(a), however, an optimally doped 20 QL sample yielded a well-defined frequency of oscillation, $\varGamma = 104.5$ T. From $\varGamma = \hbar S_F/(2\pi e)$, where $S_F = \pi k_F^2$, the Fermi wave vector, $k_F$, was extracted, and the sheet carrier density can be extracted from this and compared with the Hall effect as follows (see Ref. [25]).

As shown on the right side of Fig. 2(b), if two 2D TSS (from top and bottom surfaces) contribute, then the two surfaces will individually contribute a spin-polarized 2D carrier density of $n_{2D}^{SdH} = k_F^2/(4\pi)$. This yields a total carrier density of $2 \times n_{2D}^{SdH} = 5.0 \times 10^{12}$ /cm$^2$, which fully matches the Hall effect value of $n_{Hall} = 5.0 \times 10^{12}$ /cm$^2$ (Fig. 2(c)) and is also consistent with the previous estimations based on ARPES data (see Ref. [25]). For consistency, if we assume an elliptical bulk 3D Fermi surface as the origin of the SdH oscillations (Fig, 2(b), left side), we obtain a bulk carrier density of $n_{3D}^{SdH} \approx 6 \times 10^{18}$ /cm$^3$, which leads to conflicts with many other quantities. First of all, such a bulk state should yield an areal carrier density of $n_{3D}^{SdH} \times t \approx 12 \times 10^{12}$ /cm$^2$, which cannot be reconciled with the independently measured Hall effect, $n_{Hall} = 5.0 \times 10^{12}$ /cm$^2$. This is *unphysical* because the SdH oscillations, which capture only the high mobility channel, *can never yield an areal carrier density larger than the Hall effect*, which captures all carriers regardless of their mobilities. Also, such a bulk carrier density will put the surface Fermi level about 290 meV above the Dirac point, and thus a parallel surface carrier density of ~2.0 × 10$^{13}$ /cm$^2$ should also be present yielding a Hall carrier density of ~3.0 × 10$^{13}$ /cm$^2$, which is about six times larger than the measured value of $n_{Hall}$. Another inconsistency is that a bulk state of $6 \times 10^{18}$ /cm$^3$ should clearly show up in the $n_{Hall}$ vs *thickness* plot, Fig. 1(e). For example, at 150 QL, $n_{Hall}$ should be around $6 \times 10^{18}$ /cm$^3 \times 150$ nm $\approx 10^{14}$ /cm$^2$, which is more than an order of magnitude larger than the measured value. Altogether, we can conclude that the origin of the SdH oscillations is from the TSS, and there is negligible bulk contribution. It is worth noting that in all previous reports, the surface SdH oscillations accounted for only a small fraction of the total sheet carrier density as measured by the Hall effect.

Lastly, the decay of the oscillations with temperature and magnetic field yielded a cyclotron mass of $m^*/m_e \approx 0.13 \pm 0.01$, and quantum mobility of around ~1400 cm$^2$/Vs. The former is exactly consistent with the value that is estimated from ARPES (see Ref. [25]), and the latter is slightly lower, but fully consistent with the Hall mobility. The Berry's phase can be obtained by extrapolating linearly the Landau index, *n*, as $1/B \to 0$, which gives $n \approx 0.25$ from the lowest resolvable Landau indices of $n = 12$-16. Since this is very far from the so-called quantum limit (Landau index of 1) the extrapolated error is large, which yields inconclusive results as to the Berry's phase of this channel [9, 29]. Altogether, the analyses of the SdH oscillations are fully consistent with the surface band parameters as inferred by ARPES, and for the first time, account for the entire Hall carrier density.

With the general transport properties established above, we will now discuss the main finding in this study: the decoupling of the surface states with increasing thickness and Cu concentration. This was found by analyzing the WAL effect [2, 13, 16, 18-19], which appears as a cusp in magneto-conductance at low magnetic field (Fig. 3(a)). This magneto-conductive effect is quantitatively described by the Hikami-Larkin-Nagaoka (HLN) formula: $\Delta G(B) = \tilde{A}e^2/(2\pi h)[ln(B_\varphi/B) - \Psi(1/2 + B_\varphi/B)]$, where *h* is Planck's constant, $\tilde{A}$ is a parameter related to the number of conducting 2D channels ($\tilde{A} = 1$ for each channel), $B_\varphi$ is the de-phasing field, and $\Psi(x)$ is the digamma function[30]. The channel number, $\tilde{A}$, in undoped Bi$_2$Se$_3$ thin films is always found to be around ~1, that is one channel, independent of sample thickness[13, 16]. This implies that the two TSS in undoped Bi$_2$Se$_3$ films are coupled to the conducting bulk state and together they effectively act as one channel. Here, it is important to note that this indirect coupling in the WAL effect occurs far beyond the thickness limit (~6 QL) of direct coupling caused by wavefunction overlap between top and bottom surfaces previously observed in ARPES[20].

In Cu-doped $Bi_2Se_3$ films the behavior is quite different. Fig. 3(b) shows how the effective number of the surface states changes with increasing Cu-doping. The first thing to notice is that at doping levels above and below the critical value of 2.5 – 4.0%, the channel number is nearly constant at ~1. Consistent with other transport properties, this indicates that the bulk is conducting and the entire film effectively acts as one channel. In the critical regime, the channel number sharply increases to ~2. This coincides with the minimum in the carrier density vs. $x$ (see Fig. 1(c)). Fig. 3(c) shows $\tilde{A}$ at fixed $x$, while varying the thickness. As mentioned before, undoped $Bi_2Se_3$ yields $\tilde{A} \approx 1$ independent of thickness. However, optimally doped Cu-$Bi_2Se_3$ films exhibit qualitatively different thickness dependence. For thickness less than 10 QL, the fitting parameter $\tilde{A}$ quantizes at 1, corresponding to one channel; from 10 – 20 QL $\tilde{A}$ smoothly increased to 2, corresponding to two channels; and from 20 –150 QL $\tilde{A}$ stays constant at 2.

A simple explanation for the data shown in Fig. 3(b-c) is as follows (see Fig. 3(d) i-iv for corresponding schematic): below 10 QL, the TSS on the top and bottom surfaces can communicate (via tunneling or hopping) through the thin bulk, even if extended bulk states are suppressed, and remain strongly coupled, thus behaving as one channel. Between 10 – 20 QL, the TSS start to decouple as the communication dies out, and above 20 QL, which is too thick for inter-surface coupling to occur, the two TSS are completely decoupled from each other and behave as two isolated channels, as expected for bulk insulating systems[18, 31-32].

This process can be better described by comparing the scattering times between available states at the Fermi level and the de-phasing time in the WAL effect[18]. If the bulk has extended states at the Fermi level, surface electrons can coherently scatter in and out of the bulk states before the backscattered electrons de-phase; then, the system will behave effectively as one channel because the coherent time-reversible paths of all the transport channels are mixed up: this is the case for $Bi_2Se_3$ films with conducting bulks states as described in Fig. 3(d) i. Even if extended bulks states are absent at the Fermi level, if the opposite surfaces are within tunneling (or hopping) regime such that the inter-surface tunneling (or hopping) time is shorter than the de-phasing time, then the system should still behave as one channel: Fig. 3(d) ii. However, if available bulk states are localized at the Fermi level and the sample becomes thick enough such that the inter-surface scattering time becomes larger than the WAL de-phasing time, the two surfaces behave as two separate channels in the WAL effect because electrons on each surface complete their time-reversal paths before they scatter into opposite surfaces [18, 31-32]: Fig. 3(d) iv.

This is the first demonstration of how the TSS on opposite surfaces can become decoupled with compensation doping and thickness control. This sheds light on the distinction between the direct coupling of surface states caused by wavefunction overlapping, and the indirect coupling that is mediated by the conducting bulk states. While the direct coupling effect exists only below ~6 QL according to ARPES measurements, the indirect coupling persists beyond hundreds of QLs in the standard bulk conducting TI samples according to transport measurements. Here we have demonstrated that as this indirect coupling is eliminated by suppressing the parasitic bulk channels, the thickness dependence of the ideal TSS emerges even in transport channels. Still, the critical decoupling thickness of 10~20 QL observed in the present work is significantly bigger than the critical thickness (6 QL) observed in ARPES, and whether this is due to the difference in sensitivity between transport and ARPES probes or due to an incomplete understanding of the underlying physics is an open question and needs further studies. Altogether, this study has brought us one step closer to the realization of pure topological devices.

**Acknowledgements**

The work is supported by IAMDN of Rutgers University, the National Science Foundation (NSF DMR-0845464) and the Office of Naval Research (ONR N000141210456). We thank N. Peter Armitage and Liang Wu for discussions, Jing Chen and Weida Wu for assistance with AFM, and Yue Cao and Daniel Dessau for ARPES measurements.

**Figure captions**

**Figure 1.** Hall effect measurements as a function of Cu-doping (fixed thickness ≈ 20 QL) and thickness (fixed Cu concentration, $x ≈ 2.5$-$4.0\%$). (a) Ratio of the high field to the low field slope of $R_{xy}$. Where the ratio becomes unity, only a single carrier type contributes to the Hall effect. (b) $R_{xy}/R_{xy}(9\ T)$ vs. $B$ showing $R_{xy}$ is linear only in the optimum doping range. (c) Hall carrier density vs. $x$ showing a minimum at $x ≈ 2.5$-$4.0\%$. (d) Mobility vs. x showing the overall monotonic decrease of the mobility (black guide line) with increasing Cu doping and mobility peak (red guide line) near the optimal doping. Carrier density (e) and mobility (f) vs. thickness exhibiting nearly constant carrier density of ~$6.0 ± 2.0 × 10^{12}$ /cm$^2$ and an anomalous mobility peak near 20 QL.

**Figure 2.** SdH oscillations and comparison with Hall effect for an optimally-doped 20 QL sample ($x ≈ 2.5$–$4.0\%$). (a) $R_{xx}$ with a smooth polynomial background subtracted vs. $1/B$ taken at different temperatures. The frequency of oscillations, $\Gamma$, gives $k_{SdH}$, and therefore the carrier density $n_{SdH}$. (b) To calculate $n_{SdH}$, one must consider a 3D Fermi surface (left) versus two 2D TSS Fermi surfaces (right, one Fermi surface per TSS). The 3D Fermi surface would give an areal carrier density of at least ~$12 × 10^{12}$ /cm$^2$, while the two 2D surfaces yield $5.0 × 10^{12}$ /cm$^2$. SdH oscillations can never produce sheet carrier densities higher than the Hall effect, ruling out the bulk origin of the SdH oscillation. (c) The Hall resistance, $R_{xy}$, yields a carrier density of $n_{Hall} ≈ 5.0 × 10^{12}$ /cm$^2$, which matches the 2D SdH estimate. (d) Amplitude of the SdH oscillation vs. temperature; the fit yields a cyclotron mass of $0.13 ± 0.01$ times that of the electron mass. This agrees with the effective mass of the TSS as estimated from the ARPES spectra shown in Ref. [25].

**Figure 3.** Weak anti-localization (WAL) effect showing the thickness and doping dependent coupling of the TSS. (a) The change in conductance taken at low field showing the WAL effect. The solid lines are the raw data, and the solid squares are the fit to the HLN formula. In (b) and (c), $\tilde{A}$ shows the effective

number of 2D channels as a function of (b) composition ($x = $ Cu/Bi $\times$ 100%) for fixed thickness (20 QL) and (c) thickness for fixed composition ($x \approx 2.5 - 4.0\%$). (d) A cartoon showing top and bottom surface coupling in the WAL effect. (i) Samples with extended bulk states behave as one channel over a wide thickness range, whereas the optimally Cu-doped samples with localized bulk states exhibit (ii-iv) decoupled surface states as the film becomes thicker than an inter-surface coupling length.

**Figure 1 (single column)**

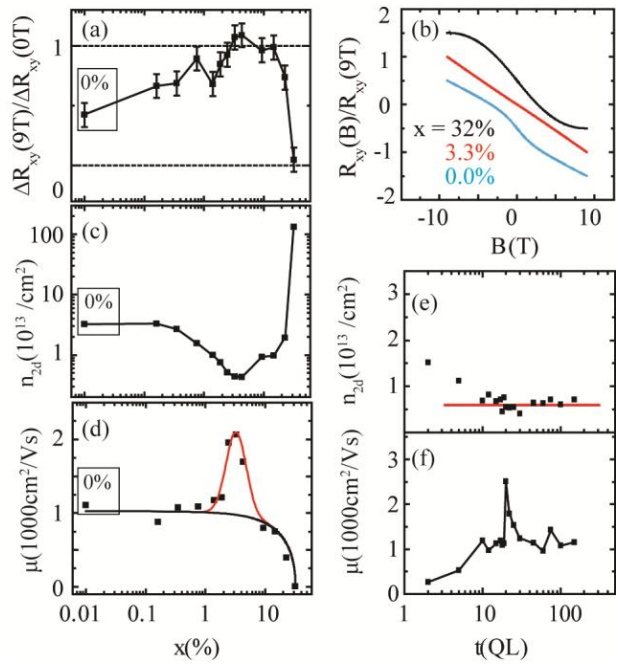

**Figure 2 (single column)**

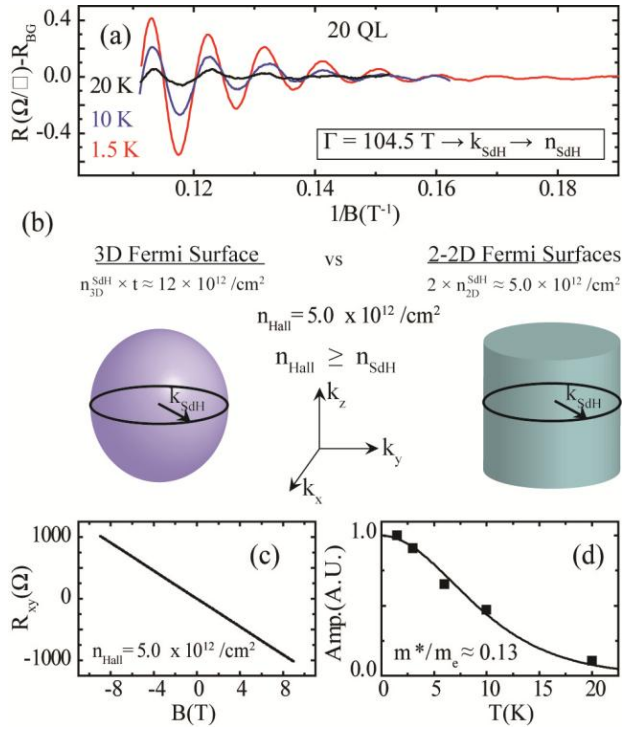

**Figure 3 (double column)**

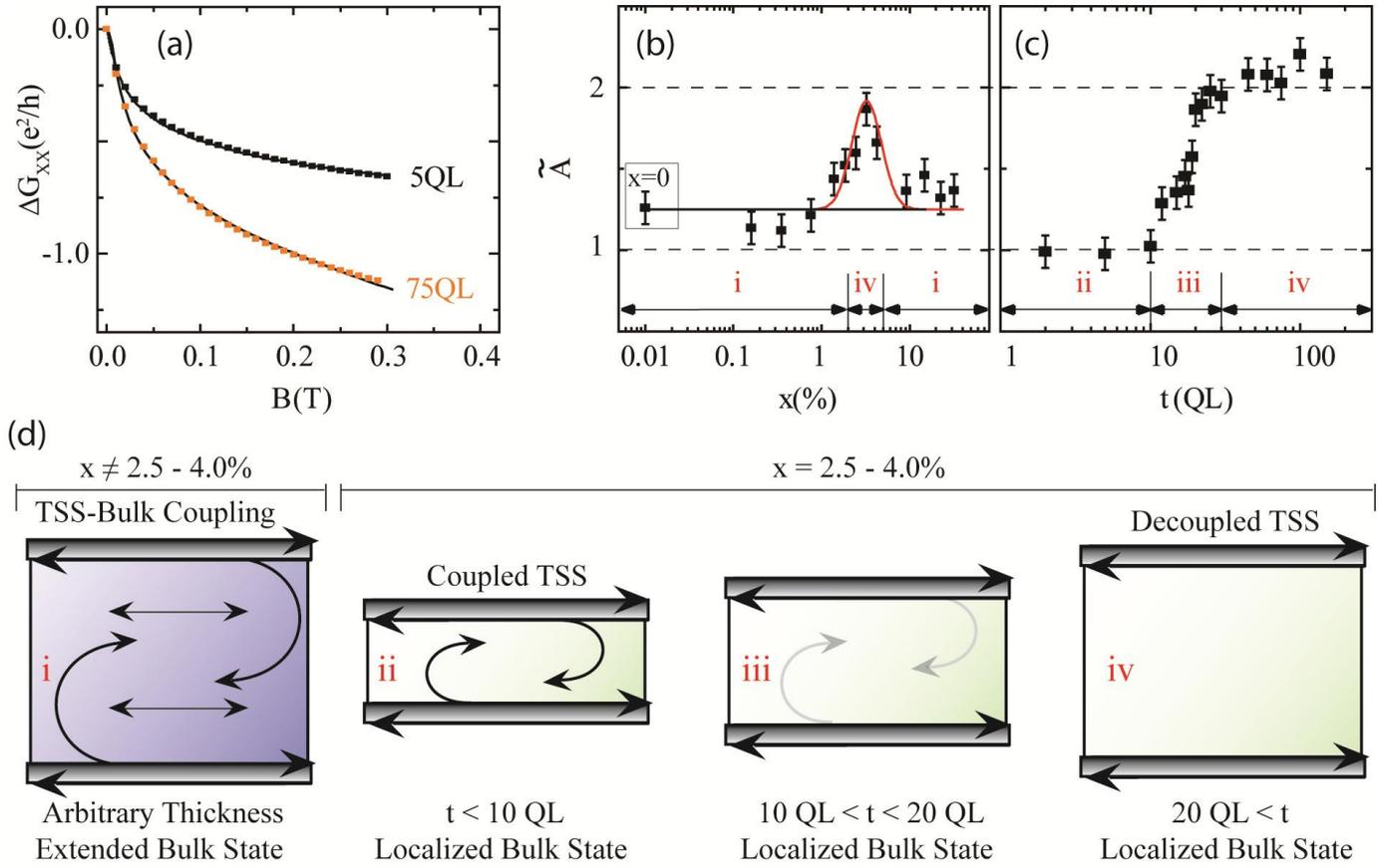

# Supplemental materials for

# Emergence of decoupled surface transport channels in bulk insulating Bi$_2$Se$_3$ thin films

Matthew Brahlek[1], Nikesh Koirala[1], Maryam Salehi[2], Namrata Bansal[3], and Seongshik Oh[1,*]

[1]Department of Physics & Astronomy, Rutgers, The State University of New Jersey, Piscataway, New Jersey 08854, U.S.A.

[2]Deparment of Materials Science and Engineering, Rutgers, The State University of New Jersey, Piscataway, New Jersey 08854, U.S.A.

[3]Department of Electrical and Computer Engineering, Rutgers, The State University of New Jersey, Piscataway, New Jersey 08854, U.S.A.

*Correspondence should be addressed to ohsean@physics.rutgers.edu

**Contents**

- **A** Methods and Materials
- **B** Compensation-Doping of Bi$_2$Se$_3$ (Fig. B 1-2)
- **C** Angle Resolved Photo-Emission Spectra Compared with Transport Data (Fig. C1, and Table C1)
- **D** Lifshitz-Kosevich Analysis of the Shubnikov-de Hass Oscillations
- **E** Origin of Suppressed Bulk Conduction (Fig. E 1-4)
    - Thickness-Dependent Band Bending
    - Impurity Band
- **F** Hall Effect and Mobility Calculation

# A: Methods and Materials

The samples used in this study were prepared by molecular beam epitaxy (SVT Associates) on $10 \times 10$ mm$^2$ Al$_2$O$_3$(0001) substrates using the standard two-step growth method for Bi$_2$Se$_3$ developed at Rutgers University[1-2] where an initial 3 QL was deposited at 135 °C, followed by remaining deposition at 300°C. The Bi, Se, and Cu fluxes were calibrated *in situ* by a quartz crystal micro-balance, and *ex situ* using Rutherford back-scattering. To track the doping level, the same Se and Bi cell temperatures were used for all samples, and by varying the temperature of the Cu cell we were able to control the doping level to within ~ ±1.0% of the target concentration.

The films produced were single crystals Bi$_2$Se$_3$ free of any Cu-based intergrowth; Fig. A1 shows structural measurements confirming this. Fig. A1(a) shows X-ray diffraction (XRD) scans for Cu doped Bi$_2$Se$_3$ at various doping levels. The only peaks detected were the standard (003n) peaks of Bi$_2$Se$_3$, and the peak of the underlying Al$_2$O$_3$ substrate (at ~41°). Similarly for reflection high energy electron diffraction (RHEED) images shown in Fig. A1(b-c), no extra peaks were observed in between the main peaks indicating no second phase existed. Both XRD and RHEED showed sharp peaks, indicating the films were of high quality single crystals that were atomically flat. This was confirmed by topographic measurements by atomic force microscopy (AFM) shown in Fig. A1(d-f).

All transport measurements were performed by attaching leads in the standard Van der Pauw geometry for a square sample, and proper averaging was used to obtain the resistance per square value. Transport measurements were preformed in an American Magnetic Inc. cryostat capable of magnetic fields up to 9 T, and down to a minimum temperature of 1.5 K. The films were briefly exposed to atmosphere during the transfer from the MBE to the liquid helium cryostat. This transfer took generally between ~3 – 5 minutes, which accounts for the sample-to-sample variation in carrier density and mobility versus thickness shown in Fig. 1(e-f) of the main text.

**Fig. A1.**

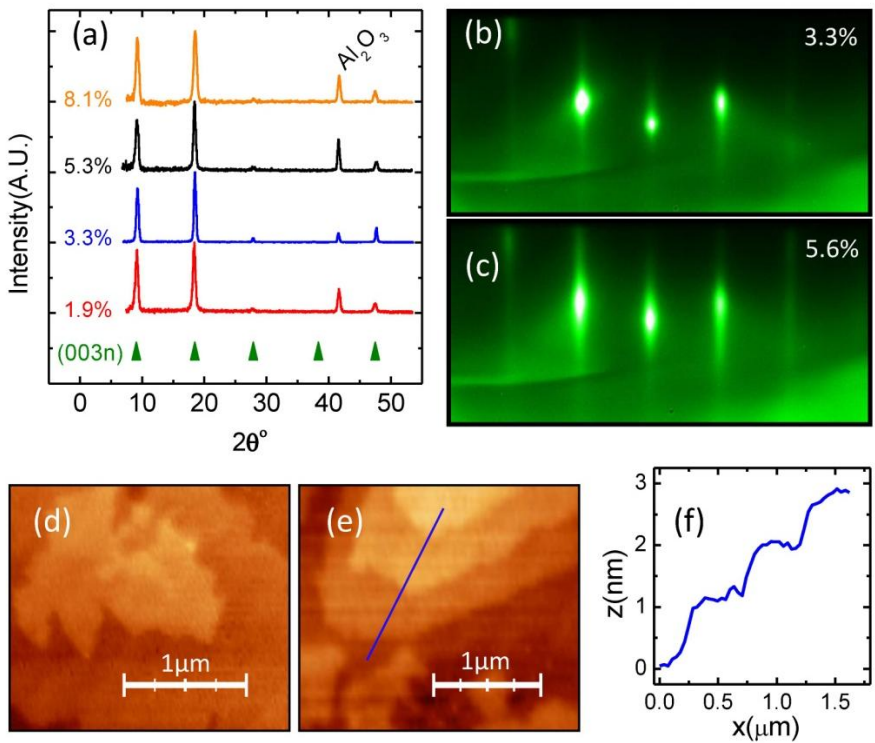

**Caption Fig. A1.** Structural measurements showing that the Cu-doped $Bi_2Se_3$ remain single phase. Cu-intergrowth would appear as extra peaks in the diffraction plots (a-c) or as clusters in the topographical images (d-e). (a) X-ray diffraction data at various values of $x$ with the thickness fixed at 20 QL. The triangles mark the (003n) peaks of $Bi_2Se_3$, and the peak at ~41° marks the $Al_2O_3$ substrate peak. (b) Reflection high energy electron diffraction images also only show peaks for $Bi_2Se_3$ with no second phase present. (d-f) Atomic force microscopy images for $x \approx 3.3\%$ show large flat terraces with feature size in excess of 1μm. The line cut in (f) corresponds to the blue line in (e), which shows the step height to be ~ 1 nm -- the height for a single quintuple layer.

## B: Compensation-Doping of $Bi_2Se_3$

Unlike traditional semiconductor systems, compensation-doping $Bi_2Se_3$ has been found to be especially tricky. The energetics of the crystal structure make it possible for the dopants to naturally sit in many different locations within the lattice. This gives rise to different electrical behavior ranging from electron donor to electron acceptor, and even neutral behavior where there is no change in the carrier density. The exact behavior depends on the specific dopant and the conditions used to grow the crystal, and therefore molecular beam epitaxy (MBE) grown films may have a different behavior compared to single crystals.

This is best illustrated by comparing the successful use of Ca to compensate-dope bulk crystals of $Bi_2Se_3$ compared to our attempts to use Ca and Zn (both ionize to 2+, and therefore should be electron acceptors) to dope MBE thin films. In the bulk crystal case, Ca was successfully used to make $Bi_2Se_3$ p-type at levels as low as 0.25% ($x$ = Ca/Bi×100%)[3]. This value is slightly larger than the value where exact charge compensation should occur; these undoped bulk crystals had a bulk electron density of ~$10^{18}$/cm$^3$, therefore the doping level for exact charge compensation is around ~0.1%, which results in a doping efficiency greater than ~ 50% (the remaining Ca atoms boost the hole carrier density). For Ca and Zn (a similar result was reported for Zn in ref. [4]) doping $Bi_2Se_3$ thin films grown by MBE no change in the carrier density was observed to a doping level beyond ~10% which results in a doping efficiency of 0%. The difference between the behavior of Ca in MBE crystals and bulk crystals likely stems from the growth temperature. For MBE the films were grown at a temperature of around ~ 300°C for roughly an hour, while the bulk crystals were grown at a much hotter temperature of ~ 800°C for several days. The higher temperature can change the growth mode and provides more thermal energy for the dopant atoms to overcome any barrier to the chemical reaction with the $Bi_2Se_3$.

As mentioned before the difference in the electrical behavior of the dopants stem from the energetic of the $Bi_2Se_3$ crystal lattice. In $Bi_2Se_3$ there are three possible positions for the dopants to reside. As illustrated in Fig. B1(a-c) the dopant atoms can replace Bi atoms and chemically bond in the lattice, they can sit interstitially in between the chemically bonded Bi-Se sites, or lastly they can intercalate in between the van der Waals bonded QL. In the first case, if a valence 2+ atom, such as Ca, Zn, or Cu, replaces Bi then it will contribute 1 hole since it replaces $Bi^{3+}$. In the remaining two cases the dopants will contribute electrons, if ionized, or may remain inert and not contribute any electrons; this depends on the exact details of the crystal growth. Our transport measurements definitively show that in the case of MBE $Bi_2Se_3$ doped by Ca, Zn, or Cu, the majority of the atoms do not contribute any electrons, and therefore most of the atoms are either interstitial or intercalated in the case of Cu doping. This is confirmed by the X-ray photoemission spectroscopy measurements shown in Fig. B2, which shows that the majority of Cu atoms in $Bi_2Se_3$ are in the $Cu^{0+}$ state and remain effectively inert (if all Cu atoms contributed charge carriers, the Hall carrier density would be over two orders of magnitude larger than observed here). Moreover, these X-ray photoelectron spectra (XPS) show that Cu-$Bi_2Se_3$ and pure Cu show nearly identical Cu 2p1/2 and 3/2 peaks, in both binding energy and line shape. This shows that the majority of the Cu sits either interstitially or intercalated in the QL gaps, and remains in an inert 0+ state working as either neutral (for low Cu content) or n-type (for high Cu content) dopants.

To see the main difference between Ca and Zn vs. Cu we can look at the difference in the carrier density versus doping level. Roughly speaking Cu only shows a small change in the carrier density compared to the amount of Cu incorporated, whereas Ca and Zn showed no change whatsoever. This shows that most of the Cu remains inactive. There are two likely mechanisms: the first, and most obvious, is Bi replacement as shown in Fig. B1(a). For this case the reason for the better success of Cu, may stem from the electronegativity: Ca and Zn (~1.00, and ~1.65) are too different from Bi (~2.02), whereas Cu, whose electronegativity is ~1.90, is a much closer match to Bi. The second scenario may be that during the growth, Cu acts mostly as a flux, facilitating the growth of $Bi_2Se_3$ with fewer defects but not playing an active role electrically; this situation accounts for the increase in mobility near the optimum doping shown in Fig. 1(d) of the main text. Overall, based on the current data we cannot say definitively the mechanism for the success of Cu doping in $Bi_2Se_3$, and this will remain a topic for further study.

In addition to the complications of the multiple locations of the dopant atoms and their different electrical behaviors, there is another complication. $Bi_2Se_3$ has both bulk and surface states, and depending on the alignment of their Fermi levels can frequently lead to band bending effects. These effects are apparent in the transport data shown in the main text. Our undoped $Bi_2Se_3$ thin-films have a bulk doping level that sets the bulk carrier density to be around ~$5 \times 10^{17}$ /cm$^2$ or less, which corresponds to $E_F \approx 15$ meV above the conduction band minimum. On the other hand, the surface carrier density of these undoped $Bi_2Se_3$ thin-films is around $1.5 \times 10^{13}$ /cm$^2$, which corresponds to a surface $E_F \approx 500$ meV above the Dirac point. Considering that the Dirac point is ~220 meV below the conduction band minimum, this implies that the undoped $Bi_2Se_3$ thin-films have downward band bending and accumulation layers on the surfaces[2]. For Cu-doped $Bi_2Se_3$, the surface Fermi level has dropped to around ~70 meV below the bottom of the conduction band. The bulk Fermi level may have also dropped a little, but in thick films it can never drop below the conduction band minimum because of the Mott criterion as discussed in section E; in other words, the bulk Fermi level is almost pinned at the bottom of the conduction band. This results in upward band bending and depletion layers on the surfaces of optimally-Cu-doped $Bi_2Se_3$ thin-films. This indicates that the main doping effect takes place at the surface, which is associated with the change in band-bending from downward (accumulation) to upward (depletion); more evidence and details on band-bending can be found in section E.

**Fig. B1.**

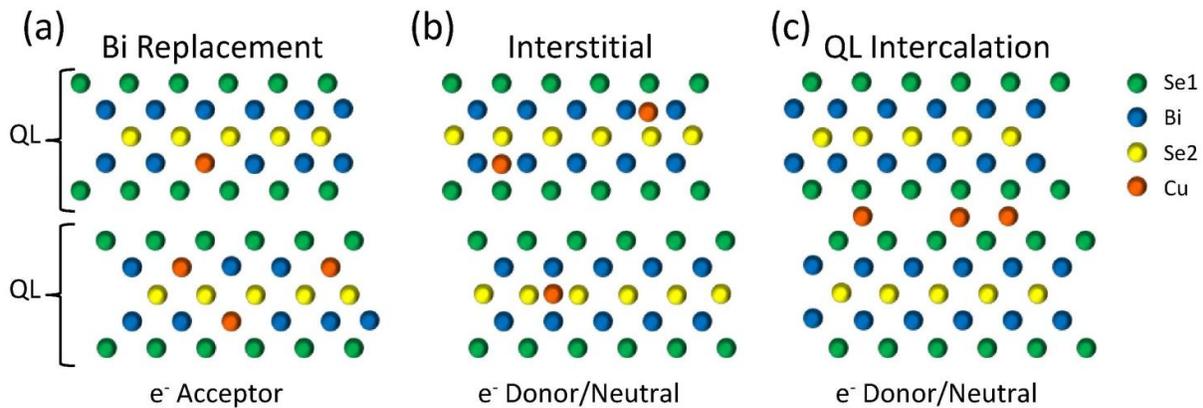

**Caption Fig. B1.** Schematic diagrams showing how doping $Bi_2Se_3$ by Cu occurs. For charge doping to occur, Cu atoms can replace Bi atoms as in (a), they can sit interstitially in (b), or they can intercalate in between the quintuple layers (QL) as in (c). In (a) Cu atoms act as electron acceptors, while in (b) and (c) they can act as electron donors. However, types (b) and (c) can act as neutral dopants depending on the microscopic details, as discussed in the text. Overall, transport and XPS measurements indicate that Cu-doping works as both p- and n-type dopants, and therefore all of the above mechanisms should be present in our MBE grown thin films.

**Fig. B2.**

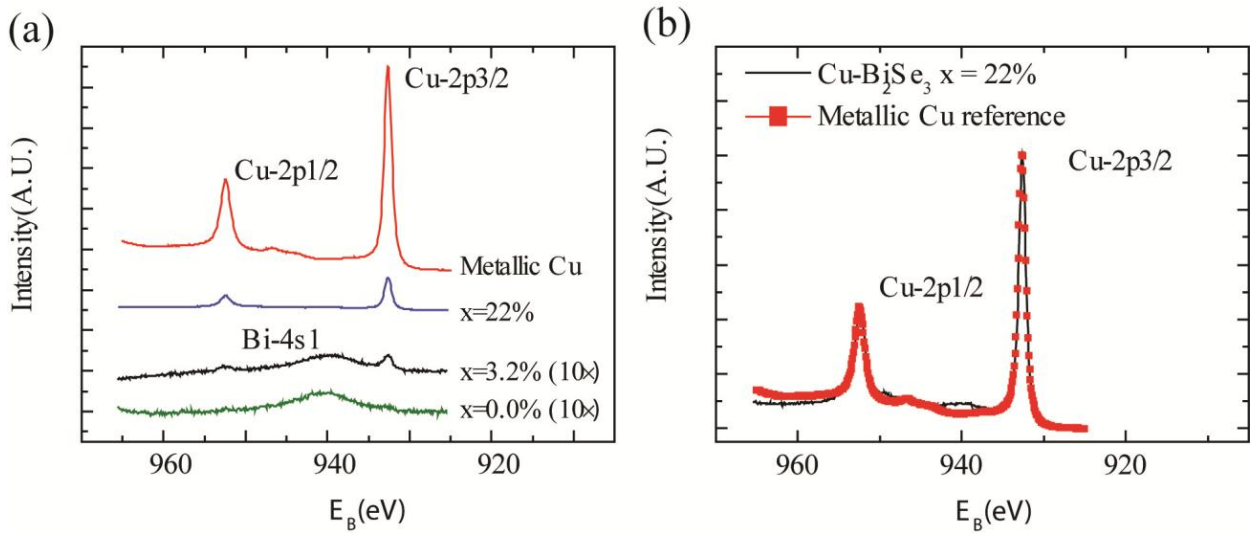

**Caption Fig. B2.** XPS data for Cu-2p3/2 and 2p1/2 for bulk Cu and Cu-$Bi_2Se_3$ with $x \approx 0 – 22\%$. (a), The Cu-peaks occur at the same binding energies (~932.6 eV and 952.5 eV respectively) for pure Cu and Cu-doped $Bi_2Se_3$. The $Bi_2Se_3$ also show a small peak in between the Cu peaks, which is the Bi-4f state (~940 eV). Binding energies are relative to C-1s, and $x \approx 0.0\%$ and 3.2% are multiplied by 10 for clarity. (b), In addition to the peaks occurring at the same binding energies, the curve shapes are nominally identical. This can be seen by superimposing the bulk Cu spectra on top of the Cu-$Bi_2Se_3$ with $x \approx 22\%$. Altogether the XPS data in (a-b) confirm the results from transport measurements (Fig. 1 main text) that show the majority of the Cu remains in an unionized state.

## C: Angle Resolved Photo-Emission Spectra Compared with Transport Data

The clearest studies of the TSS have been done using angle resolved photo-emission spectroscopy (ARPES). Therefore, it is best to use their direct measurements of the band-structure as a guide to understand our transport data. To minimize ambiguous estimations, we made use of an empirical curve fit to the surface dispersion relations to estimate the parameters that we compare with transport data. As detailed elsewhere[5], an isotropic dispersion for the TSS can be fit to ARPES spectra that yields $E_{ave}(k) = 1.9k + 12.6k^2 + 2300k^6$, where $k$ is in units of 1/Å and $E$ is in units of eV. As shown in Fig. C1, this function is overlaid on an typical ARPES spectrum taken from ref. [6], which was chosen for its clarity. The empirical function fits equally well all ARPES spectra taken for bulk crystals, and thin films (both undoped and Cu-doped). Therefore, from this we calculate and then compare measured transport properties with what we expect from the empirical band structure.

One quantity that is typically measured by transport is the carrier density per area, $n_{Hall}$. This is done by measuring the transverse Hall resistance, $R_{xy}$, which is then related to the carrier density by $R_{xy} = B/(en_{Hall})$, where $B$ is the magnetic field, and $e$ is the electron charge. From this we can directly calculate the Fermi wave vector to be, $k_F = 0.056$ /Å, from $n_{Hall} = 2k_F^2/(4\pi)$, where the two in the numerator is from the fact that there are TSS on both the top and bottom surfaces. From here, we can use the empirical dispersion curve to calculate the Fermi energy to be $E_F \approx 146$ meV above the Dirac point, and about ~69 ± 15 meV (depending on the distance from the conduction band minimum and the Dirac point, which we take to be 215 ± 15 meV) below the conduction band bottom. At this energy, the Fermi velocity is $v_F(k) = dE/d(\hbar k) \approx 5.0 \times 10^5$ m/s, and the cyclotron mass is $m^*(k)/m_e = \hbar k/(v_F m_e) \approx 0.13 \pm 0.01$. As shown in Table C1, these values can be compared to the data obtained by Hall effect measurements and SdH oscillations.

**Fig. C1.**

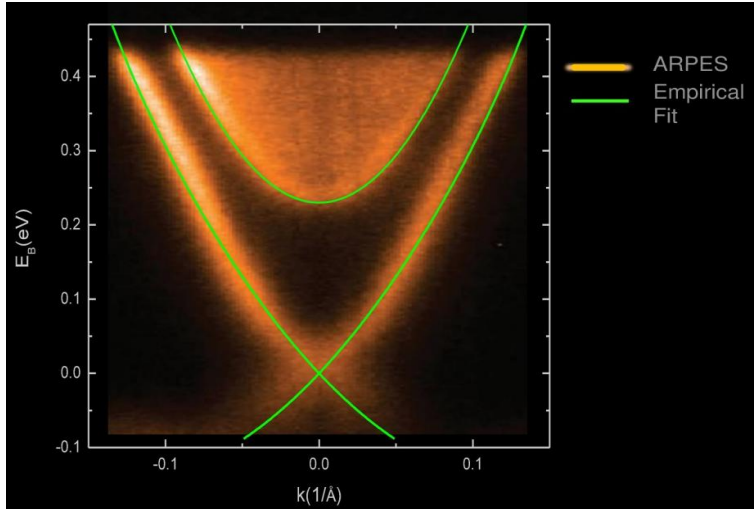

**Caption Fig. C1.** An ARPES spectrum taken from ref. [6] overlaid by the empirical functions, $E_{ave}(k) = 1.9k + 12.6k^2 + 2300k^6$ for the TSS and $E_{CB}(k) \approx \hbar^2 k^2/(2m^*) + E_{DP}$ for the conduction band with $m^*/m \approx 0.15$ and $E_{DP} \approx 230$ meV. Parameters obtained from this function can be directly compared to transport data (see Table SI -C1).

**Table C1.**

| Measurement | $k_F$ (1/Å) | $v_F$ ($10^5$ m/s) | $n_{2D}$ ($10^{12}$/cm$^2$) | $m^*/m$ | $E_F$ (meV) |
|---|---|---|---|---|---|
| **Hall Effect** | 0.056 | - | 5.0 | - | - |
| **SdH Oscillations** | 0.056 | 5.0 | 5.0 | 0.13±0.01 | - |
| **ARPES (@$k_{F,Hall}$)** | 0.056 | 5.0 | - | 0.13±0.01 | 146 |

**Caption Table C1.** Comparison of data obtained by the Hall effect, SdH oscillations, and values calculated via an empirical curve fitted to an ARPES spectrum. From this we see that all three techniques agree that the measured transport properties emanate from the TSS.

## D: Lifshitz-Kosevich Analysis of the Shubnikov-de Hass Oscillations

Shubnikov-de Hass (SdH) oscillations appear when a metal is subjected to a sufficiently strong magnetic field (such that $\mu \times B > 1$, where $\mu$ is the electron mobility, and $B$ is the magnetic field. $\mu \times B \approx 2.5$ for our films.). The magnetic field collapses the continuous energy bands into a spectrum of individual energy levels. As the magnetic field increases, the energy level spacing increases. Therefore, increasing $B$ pushes these energy levels across $E_F$, and thus gives rise to the resistance oscillating with increasing magnetic field [7].

A standard Lifshitz-Kosevich [7] (LK) analysis was carried out as follows. The SdH oscillations of the resistance follows the LK formula $R_{SdH} = R_T R_D cos[2\pi(\Gamma/B+1/2+\beta)]$, where $\Gamma$ is the period of oscillation, $2\pi\beta$ is the Berry phase. The first prefactor is given by $R_T = (\alpha T/B)/sinh(\alpha T/B)$, with $\alpha = 2\pi^2 m^* k_B/(\hbar e)$, where $m^*$ is the cyclotron mass, $k_B$ is Boltzmann's constant, $T$ is the temperature. The second prefactor is the Dingle factor $R_D = exp[-\pi m^*/(eB\tau_D)]$, where $\tau_D$ is the Dingle scattering time, which is related to the quantum mobility by $\mu = e\tau_D/m^*$.

By measuring $\Delta R_{xx}$ vs. $1/B$ (Fig. 2(a) in the main text) at different temperatures, we can extract the cyclotron mass. As shown in Fig. 2(d) of the main text, at a fixed magnetic field the amplitude of the oscillations decays with increasing temperature. This isolates the $R_T$ term in the LK formula: by fitting $R_T$ vs. $T$ at fixed magnetic fields, we can extract the cyclotron mass. Once $m^*$ is found, $R_T$ is determined, and then the Dingle factor, $R_D$, can be extracted by fitting the oscillations at fixed temperature to the LK formula, and then we can calculate $\tau_D$.

One standard use of SdH oscillations is to separate a 3D Fermi surface from a 2D Fermi surface. This is done by measuring the resistance versus magnetic field that is applied at an angle relative to the film's surface: if the oscillation is from a 3D Fermi surface, it should be roughly angle-independent, whereas if it is from a 2D Fermi surface, it should depend only on the surface normal component of the applied magnetic field. In the past, identification of the 2D Fermi surfaces by SdH oscillations has been used as direct evidence that the transport originates from the TSS. However, this is not necessarily true. In the case of a topologically trivial high mobility thin-films, the mean free path can be many times the film thickness. The transport signature of such a film would appear as if it was 2D from SdH oscillations, even if the transport does not originate from any surface states. An even more serious complication has been pointed out recently in undoped $Bi_2Se_3$ bulk crystals. Because of the intrinsic 2D nature of the $Bi_2Se_3$ crystal structure, originating from the weak van der Waals bonding between quintuple layers, even the bulk $Bi_2Se_3$ state can exhibit 2D behavior in SdH oscillations[8]. What this shows is that the 2D signature observed by angle dependent SdH oscillations is less stringent than showing the transport is thickness independent as shown here, which is possible only if the conduction occurs through the surface states with negligible bulk contribution.

Lastly, we show another set of SdH oscillations for a 30 QL sample. The data shown in Fig. D1 are consistent with the data shown in the main text for the 20 QL sample. However the amplitude is lower, and therefore due to the space limits of the main text, we chose to include this data here. From these oscillations the frequency is slightly higher than the 20 QL, at 122.2T, and the effective mass is also slightly larger at $m^*/m_e \approx 0.14 \pm 0.01$, but the deviation is within the error bar, and therefore we cannot correlate the larger effective mass with the slightly higher Fermi level. Overall, for all samples with SdH oscillations that were prominent enough to allow a full Lifshitz-Kosevich analysis, all the data consistently pointed to the channel emanating from the topological surface states.

**Fig. D1.**

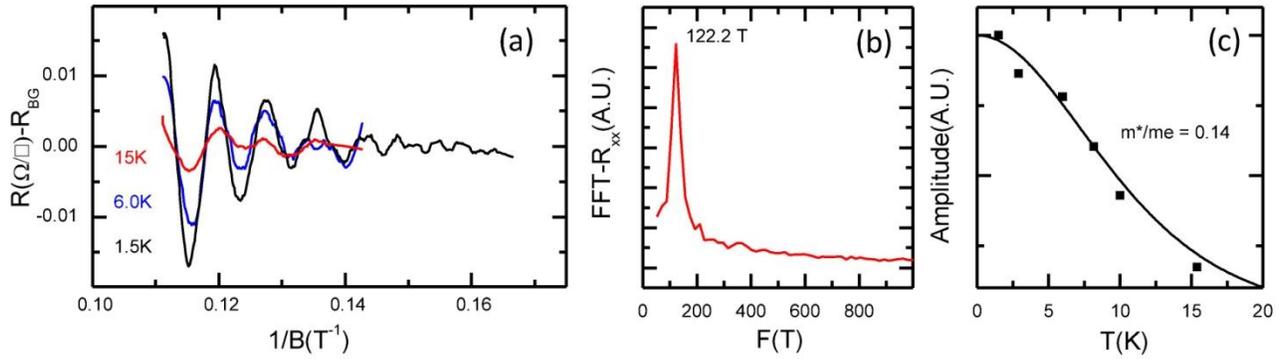

**Caption Fig. D1.** SdH oscillations for 30 QL sample. This sample showed similar oscillations to the 20 QL sample in main text. However, the amplitude is lower in magnitude. The data for this 30 QL is consistent with 20 QL in the main text, and is fully consistent with the oscillations coming from the TSS.

## E: Origin of Suppressed Bulk Conduction

As discussed in the main text, all transport measurements coherently indicate that at low temperature only the topological surface states contribute to the conduction. Based on the resistance versus temperature data shown in Fig. E1, we discuss two scenarios for the observed suppression of the bulk conduction.

**Fig. E1.**

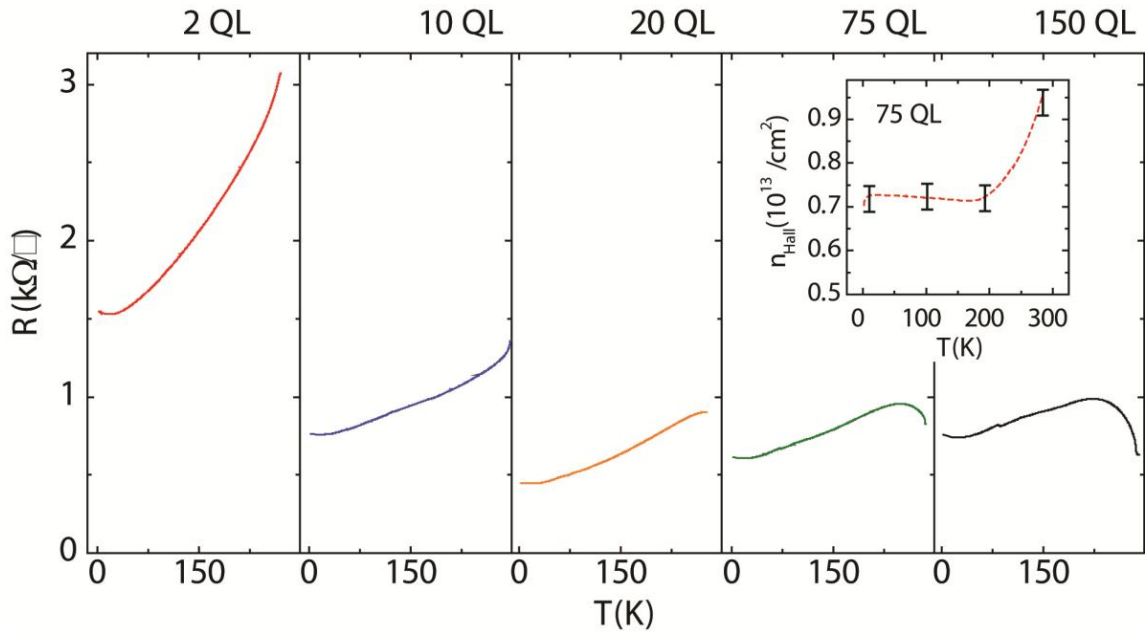

**Caption Fig. E1.** Resistance vs. temperature for $t$ = 2, 10, 20, 75, and 150 QL (1 QL ≈ 0.95 nm). The hump at high temperature (250 - 300 K) is associated with the bulk-carriers freezing-out. This can be seen in the inset, which shows $n_{Hall}$ vs. $T$ for a 75 QL film

## Thickness-Dependent Band Bending

The major problem in $Bi_2Se_3$ and all topological insulators is the existence of a bulk channel in parallel with the TSS. Much work has been done in bulk crystals and thin-films to eliminate the bulk channel (when we refer to the bulk channel, we mean this to incorporate bulk conduction plus any bulk related 2DEG—bulk is anything except the TSS)[9-14]. To date, however, the lowest volume carrier density reported is ~$10^{16}$/$cm^3$, and it was found that these crystals were dominated by bulk conductance[15]. The fact that the bulk of $Bi_2Se_3$ remained conducting even at a bulk carrier density of ~$10^{16}$/$cm^3$ is not surprising; this can be understood by considering the Mott criterion, which gives the critical dopant density where a metal-to-insulator transition occurs[16]. This critical value occurs when the mean dopant spacing is of the same order as the effective Bohr radius of the defect states. In a crystal with a dielectric constant $\varepsilon$ ($\varepsilon \approx 113$ for $Bi_2Se_3$[17]), the effective Bohr radius is given by $a_B = \varepsilon/(m*/m) \times 0.05$ nm ≈ 40 nm, where $m* \approx 0.15m$ is the effective mass and $m$ is the electron mass [18]. Then the critical dopant density, $N_{CD}$, is given by $a_B N_{CD}^{1/3} \approx 0.26$. From this we get $N_{CD} = (0.26/a_B)^3 \approx 3 \times 10^{14}$/$cm^3$. Therefore, for bulk crystals unless the dopant density is below ~$3 \times 10^{14}$/$cm^3$, their bulk states will be always metallic with $E_F$ pinned to the conduction band minimum. A dopant density of order ~$10^{14}$/$cm^3$ is achievable in semiconductors such as Si and GaAs; however, since the chemical bonding in $Bi_2Se_3$ is much weaker, it seems to be thermodynamically impossible for a defect density to reach such a low level in bulk crystals of Bi-based TI materials. However, in $Bi_2Se_3$ there is significant band-bending due to a large mismatch in the surface and bulk Fermi levels (before equilibrium), which we argue below can be employed to lower $E_F$ into the bulk band gap in thin-films.

Band-bending can be qualitatively understood by knowing the surface carrier density, $n_{ss}$, along with the knowledge that the Mott-criterion fixes $E_F$ in the bulk to be close to the conduction band minimum. In the main text we showed that if the top and bottom TSS carry a total density of $n_{ss} = 1.0 \times 10^{13}$/$cm^2$, then $E_F$ at the surface exactly touches the bottom of the conduction band. This then implies that if $n_{ss} = 1.0 \times 10^{13}$/$cm^2$, band-bending is absent, and the conduction band minimum is at the same level from the surface into the bulk (see Fig. E2(a)). If $n_{ss} > 1.0 \times 10^{13}$/$cm^2$, which is the case for undoped $Bi_2Se_3$ where $n_{ss}$ is typically ~$3.0 \times 10^{13}$/$cm^2$ (see ref. [2]), then the bands must be bent downward close to the surfaces (see Fig. E2(b)). This gives rise to an accumulation region near the surface, which has been shown to harbor a quantum-well based 2DEG[2, 6]. However, for optimally Cu-doped $Bi_2Se_3$ thin-films, $n_{ss} < 1.0 \times 10^{13}$/$cm^2$ ($n_{ss} = 5.0 \times 10^{12}$/$cm^2$, more specifically); this then implies that the bands bend upward close to the surfaces, which gives rise to a depletion region near the surface (see Fig. E2(c)); such a state cannot have a non-topological 2DEG. Thus in optimally Cu-doped $Bi_2Se_3$ the direction of the surface band-bending is upward with a depletion layer formed at the surface.

To describe the depletion region formed at the surface of the Cu-doped $Bi_2Se_3$ quantitatively, one needs to solve the Poisson equation self-consistently, but to capture a rough physical picture we can, to first approximation, solve it exactly by assuming the bulk dopants are distributed uniformly throughout the bulk. The Poisson equation is

$$\nabla^2 V(z) = -\frac{e^2 N_{BS}}{\varepsilon \varepsilon_0}$$

where $V$ is the potential energy as a function of distance from the surface, $z$, $e$ is the electron charge, and $N_{BS}$ is the bulk dopant density. Subjecting this equation to the boundary conditions of $V(z = z_d) = 0$, and $V(z = 0) = \Delta V$, where $\Delta V$ is the energy difference between the bands deep in the bulk and at the surface, and $z_d$ is the distance over which the band-bending occurs in the bulk limit of $t \gg z_d$ (see Fig. E3). The solution to Poisson's equation for one surface is then

$$V(z) = \frac{e^2 N_{BS}}{2\varepsilon\varepsilon_0}(z - z_d)^2, (z < z_d)$$
$$V(z) = 0, (z > z_d)$$

where the boundary conditions require that $\Delta V = e^2 z_d^2 N_{BS}/(2\varepsilon\varepsilon_0)$ and the electric field vanishes beyond $z = z_d$.

One thing we can note from this is that in the thin-film limit ($t < 2 \times z_d$), $z_d$ is limited to half the film thickness. Therefore when $t < 2 \times z_d$, $\Delta V$ must decrease accordingly, and hence as the thickness decreases, $\Delta V \propto t^2 \to 0$, which implies that the band-bending is too energetically costly to be sustained and vanishes as the films are made thinner (see Fig. E3(a-c)). To estimate a lower bound on $z_d$, we can use the bulk dopant density from our undoped Bi$_2$Se$_3$ films[2], whose estimated upper bound is $N_{BS} \approx 5 \times 10^{17}$ /cm$^3$, and since the Cu atoms lower the bulk donor density, we can take $N_{BS}$ in our Cu-doped films to be $< 5 \times 10^{17}$ /cm$^3$. Assuming that the surface Fermi level is ~ 70 meV below the conduction band minimum as estimated above in section B, and that $E_F$ in the bulk at most touches the conduction band minimum (based on the Mott criteria), then we can take $\Delta V$ to be ~70 meV (see Fig. E3(a)). Based on these values, the formula $\Delta V = e^2 z_d^2 N_{BS}/(2\varepsilon\varepsilon_0)$ then leads to $z_d$ larger than 40 nm (see ref. [18] where $z_d$ is estimated to be of similar magnitude). Since at the surface $E_F$ is pinned at ~70 meV below the conduction band minimum (which is set by the surface defect density), $E_F$ must drop below the conduction band in the bulk as the film becomes thinner than $2 \times z_d$, which is an order of ~100 nm (see Fig E4). The effect is a true topological phase developing in the thin-film limit ($t \lesssim 100$ nm).

The band-bending as a function of thickness provides a natural explanation for the resistance versus temperature data shown in Fig. E1. At high temperatures a hump develops for $t > 30$ QL; the size of the hump increased steadily with thickness. When the films are significantly thinner than twice the depletion length, the bands are almost flat, and the bulk $E_F$ is as much as ~70 meV below the conduction band minimum; therefore, the bulk carriers are frozen all the way up to room temperature ($k_B T \approx 26$ meV). As the films become thicker, the band-bending develops, and the gap between the bulk $E_F$ and the conduction band minimum ($E_{CB,min}$) gradually decreases. Then, activated bulk carriers can contribute to the conductance near room temperature and freeze out only at low temperatures when $k_B T \ll E_{CB,min} - E_F$, which gives rise to a hump developing in $R$ vs. $T$; this is qualitatively consistent with what we observed in Fig. E1.

**Fig. E2.**

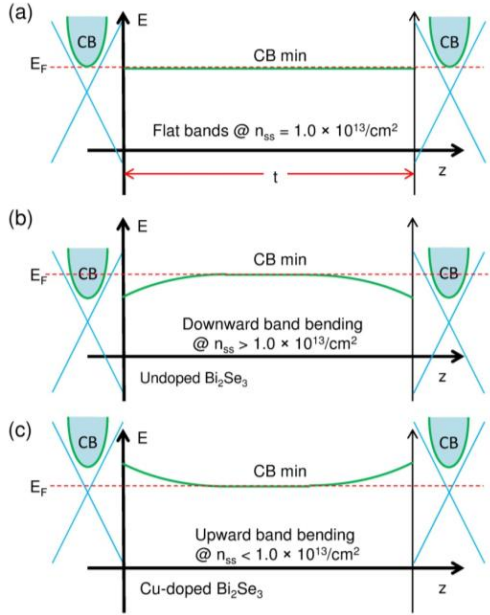

**Caption Fig. E2.** Band-bending dependence on the TSS carrier density: $n_{ss}$ represents the total carrier density from top and bottom TSS. So long as the bulk donor density remains larger than ~$3 \times 10^{14}$ /cm$^3$, the Mott-criterion requires that $E_F$ remains pinned to the conduction band (CB) minimum deep in the bulk. (a) A surface carrier density of $n_{ss} = 1.0 \times 10^{13}$ /cm$^2$ fixes $E_F$ at the CB bottom on the surface, and therefore the bands must be flat. (b) If $n_{ss} > 1.0 \times 10^{13}$ /cm$^2$, then bands must bend downward. This is the case for undoped Bi$_2$Se$_3$. (c) If $n_{ss} < 1.0 \times 10^{13}$ /cm$^2$, then bands must bend upward. This is the case for optimally Cu-doped Bi$_2$Se$_3$.

**Fig. E3.**

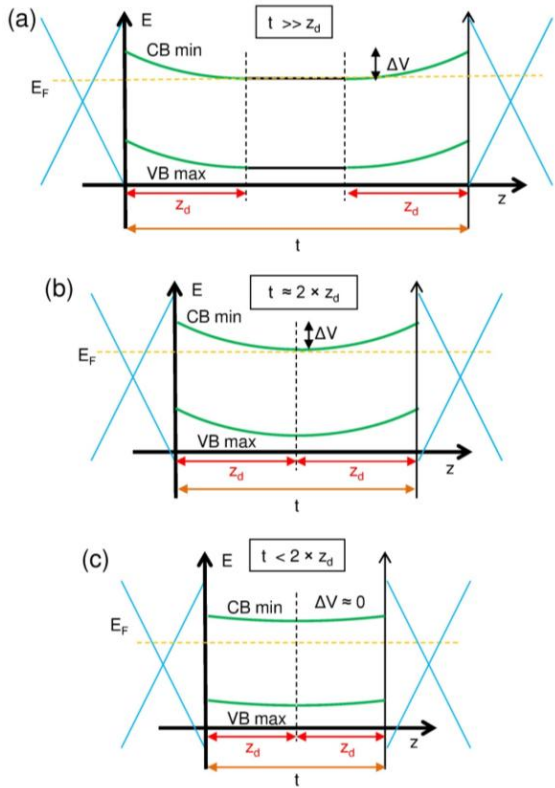

**Caption Fig. E3.** This illustrates how band bending behaves as the films are made thinner. In (**a**) $t \gg z_d$, so the film behaves as if it were a bulk sample. In (**b**) $t \approx 2 \times z_d$, and there still exists significant band-bending. In the limit of $t < 2 \times z_d$, as shown in (**c**), $z_d$ is limited by the film thickness, and hence the level of band-bending, $\Delta V$, diminishes.

**Fig. E4.**

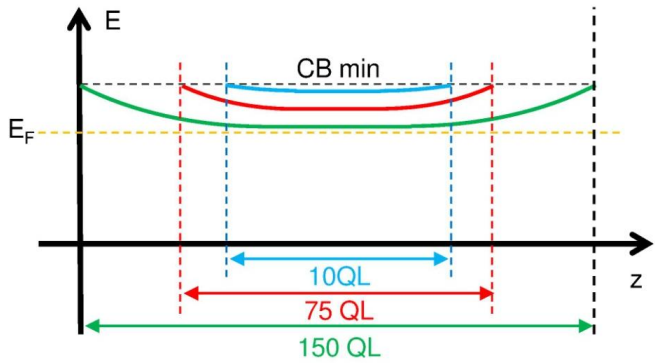

**Caption Fig. E4.** As the films are made thicker, the conduction band minimum as a function of distance from the surface dips down closer to the $E_F$.

## Impurity Band

The existence of impurity band in doped $Bi_2Se_3$ and other topological materials has been used to explain many $R$ vs. $T$ curves of low bulk conductance samples[9, 11, 19]. This is typically seen as an increase in the resistance with decreasing temperature. Therefore, despite the fact that the band-bending picture given above satisfactorily explains the data, it is appropriate to consider the impurity band effect as another scenario for the data shown in Fig. E1.

In doped materials, as the doping density increases, the mean separation between dopants decreases and the wave functions of the electrons around the dopants will begin to overlap, and may eventually form a continuous impurity band in the bulk energy gap. If the impurity band is formed near or within the conduction band, then it goes back to the band bending scenario as discussed above, but if it is formed in the middle of the bulk band gap, the band bending picture is no longer valid. Due to the random location of dopant atoms, the mobility of impurity bands are, in general, very low, and at low temperatures the impurity band may be either metallic or insulating depending on the Ioffe-Regel criterion. According to the Ioffe-Regel criterion, a band will remain metallic only if the product of the Fermi wave vector, $k_F$, and the mean-free path, $l$, is larger than unity[20]. To estimate $k_F l$, we can take, as an upper bound, $\mu_{IB} < 50$ cm$^2$/Vs [9], and $n_{3D,IB} < 5 \times 10^{17}$/cm$^3$, then $l = v_F \tau = (\hbar \mu_{IB}/e)(3\pi^2 n_{3D,IB})^{1/3}$, and $k_F = (3\pi^2 n_{3D,IB})^{1/3}$, where the Fermi velocity is given by $v_F = \hbar k_F/m^*$, and the relaxation time is given by $\tau = m^*\mu/e$. Then, for the impurity band, if it exists, we finally get $k_F l < 0.2$. Since $k_F l$ is significantly smaller than unity, this impurity band should be insulating with the carriers freezing out at low temperature. This can also explain the carrier freeze-out and the hump feature in *R* vs. *T* as observed in Fig. E1, even in the absence of the band bending effect.

If this low mobility bulk impurity band is the origin of the suppressed bulk conduction, then the inter-surface coupling will occur through hopping via the bulk states instead of direct inter-surface tunneling as would be the case for the strong insulating bulk state developed by the thickness-dependent band bending effect as discussed above. In the impurity band scenario, if the surface-bulk-surface hopping time becomes larger than the de-phasing time of the WAL effect on each surface, the WAL effect will exhibit two channels as observed here. In other words, both the impurity band picture and the thickness-dependent band bending picture can explain our data satisfactorily. In order to pinpoint the origin behind the suppression of the bulk conduction, more detailed studies will be needed.

## F: Hall Effect and Mobility Calculation

From measurement of the transverse magneto-resistance, also called the Hall resistance $R_{xy}$, and the longitudinal resistance $R_{xx}$, one can calculate the number of electrons per unit area, $n_{Hall}$, and the electron mobility, $\mu$. The classical Hall formula relates $R_{xy}$ to $n_{Hall}$ by the following formula: $R_{xy} = B/(en_{Hall})$, where *B* is the magnetic field, and *e* is the electron charge. From measurement of $R_{xx}$ the sheet conductance can be calculated and then related to the free electron mobility as $\sigma = e\mu n_{Hall}$, where $\sigma$ stands for the sheet conductance. One last thing to note is that the Hall effect yields the total electrons per area, which has the units of 1/*Area* and this should not be taken necessarily as the carriers from the surface states. Simply put, it is the number of carriers per area.